\documentclass[]{interact}

\usepackage{epstopdf}
\usepackage[caption=false]{subfig}
\usepackage[numbers,sort&compress]{natbib}
\bibpunct[, ]{[}{]}{,}{n}{,}{,}

\makeatletter
\def\NAT@def@citea{\def\@citea{\NAT@separator}}
\makeatother

\usepackage{amsmath,amssymb,bm}
\usepackage{graphicx}
\usepackage{xcolor}
\usepackage{url}

\newcommand{\draftfigure}[4][0.82\linewidth]{%
\begin{figure}[t]
\centering
\IfFileExists{#2}{\includegraphics[width=#1]{#2}}{%
\fbox{\parbox[c][4.3cm][c]{0.88\linewidth}{\centering
\textbf{Figure placeholder}\\[1mm]
}}}
\caption{#3}
\label{#4}
\end{figure}}

\begin{document}

\title{When chaos meets thermodynamics: from Fermi acceleration to thermalisation}

\author{
\name{Edson Denis Leonel\thanks{CONTACT Edson Denis Leonel. Email: edson-denis.leonel@unesp.br}}
\affil{Departamento de F\'isica, Universidade Estadual Paulista (UNESP), Av. 24A, 1515 - CEP 13.506-900, Rio Claro, SP, Brazil}
}

\maketitle

\begin{abstract}
A particle colliding with a moving wall provides one of the simplest mechanical settings in which a small system can exchange energy with its surroundings. In a chaotic time-dependent billiard, repeated elastic impacts may produce Fermi acceleration, with the characteristic kinetic energy growing without bound. This result poses an instructive thermodynamic puzzle: a physical wall maintained at finite temperature should act as a reservoir and drive a dilute gas toward a stationary temperature rather than heat it indefinitely. We use this puzzle to connect nonlinear dynamics with elementary thermodynamics. Under suitable decorrelation assumptions, collision sequences generate effective diffusion in velocity space and the square-root growth associated with normal diffusion. We then introduce two ingredients absent from the conservative accelerator: phase randomisation, which simplifies the statistics of energy exchange, and dissipation, which transfers energy back to unresolved degrees of freedom of the boundary. Their competition replaces unbounded acceleration by a stationary kinetic-energy scale. A recurrence for the second velocity moment yields a stationary kinetic-temperature scale
$T_\infty=(mA^2/8k_B)(1+\gamma)/(1-\gamma)$, where $A$ measures the boundary forcing and $\gamma$ is a restitution coefficient. We compare relaxation in collision number with a simple Fourier-inspired heat-transfer model in physical time, emphasising that the two clocks are not interchangeable during transients. The resulting picture shows how fluctuations, dissipation, coarse graining and relaxation can produce an effective thermal state from chaotic dynamics while also clarifying the limits of the thermodynamic analogy.
\end{abstract}

\begin{keywords}
Fermi acceleration; chaotic billiards; thermalisation; heat transfer; nonequilibrium statistical physics; stochastic dynamics; dissipation
\end{keywords}

\textbf{Outline of the paper.}
We begin with the collision mechanics behind Fermi acceleration, then introduce dissipation and decorrelation to explain the emergence of a stationary kinetic-energy scale. The second half contrasts impact-to-impact relaxation with a Fourier-inspired macroscopic model, distinguishes kinetic stationarity from equilibrium thermalisation, and discusses the limitations and possible extensions of the description.

\tableofcontents

\section{Introduction: can a moving wall heat forever?}

A collision with a stationary hard wall is dynamically uneventful as far as kinetic energy is concerned. The normal component of the velocity reverses, the tangential component is preserved, and the speed remains unchanged. Once the wall moves, however, the situation becomes qualitatively different. An elastic reflection in the instantaneous rest frame of the boundary is no longer elastic in the laboratory frame, and the particle may emerge from the collision either faster or slower than it arrived. This elementary observation is the starting point of Fermi's celebrated acceleration mechanism, originally proposed in the context of cosmic rays interacting with moving magnetic structures \cite{Fermi1949}.

The same idea can be realised deterministically by confining a particle within moving boundaries. Time-dependent billiards are especially clean models because the particle moves freely between impacts, so the boundary is the sole agent of energy exchange. Standard texts on billiard dynamics and Hamiltonian chaos provide the static background \cite{ChernovMarkarian2006,LichtenbergLieberman1992}. Once the boundary moves, however, the dynamics can become much richer. Chaotic collision sequences, or stochastic transport channels created by the time dependence itself, can lead to diffusion in velocity space and, under suitable conditions, to unbounded energy growth \cite{Loskutov2000,Lenz2008,Gelfreich2012,Itin2012}. This phenomenon is broadly referred to as Fermi acceleration.

The language of diffusion is useful because it separates a common misconception from the actual mechanism. Fermi acceleration does not require every collision to increase the energy; a typical trajectory contains both gains and losses. What matters is the cumulative statistics of many kicks. When successive collision phases become sufficiently irregular, the mean increment may be small while its variance continues to grow. To leading order, the velocity then acquires a random-walk character and a characteristic speed grows as the square root of the number of impacts. More systematic stochastic descriptions can be formulated in terms of transition probabilities or Fokker-Planck-type equations \cite{Karlis2012}. Correlations, resonances and phase-space structures can nevertheless alter this simple picture and produce anomalous transport or faster asymptotic growth \cite{Gelfreich2012,Itin2012,Hansen2018}.

This dynamical picture contains a thermodynamic puzzle. Suppose that the moving boundary is reinterpreted as a wall maintained at a finite temperature $T_b$. For a genuinely thermalising wall, an ensemble colder than the wall should gain energy from it. If the gas is hotter, the net energy flow should reverse. In either case a finite reservoir temperature suggests relaxation toward a stationary state. A conservative chaotic billiard, by contrast, may continue to pump energy into the particles without bound. If we translate kinetic energy into an effective temperature, the model predicts $T\rightarrow\infty$.

The apparent contradiction is not a failure of either chaos theory or thermodynamics. It exposes what the purely mechanical moving-wall model leaves out. A thermal reservoir is not merely a source of fluctuations; it must also provide a channel through which the system can return energy to its surroundings. Effective descriptions realise these two roles in different ways: random forcing and friction in Langevin models, stochastic collisions in Andersen dynamics, deterministic feedback in Nos\'e-Hoover thermostats, or probabilistic gas-surface kernels in kinetic theory \cite{Kubo1966,Andersen1980,Nose1984,Hoover1985,CookFeres2012,WuStruchtrup2017}. The microscopic implementations differ substantially. For the present purpose, the common idea is an effective competition between fluctuating energy exchange and dissipation.

Time-dependent dissipative billiards provide a particularly transparent arena in which to see this balance develop. In the model discussed in this paper, the moving boundary generates stochastic-like kicks while an inelastic collision law removes a fraction of the normal relative kinetic energy. For weak dissipation the early dynamics still resembles the conservative accelerator, but the growth eventually crosses over to a stationary velocity scale \cite{OliveiraLeonel2010,LeonelBunimovich2010}. Under a decorrelated-impact approximation, the second moment of the velocity obeys a linear recurrence, making the approach to stationarity analytically tractable. The same model can be read in thermodynamic language: the stationary second moment defines a kinetic temperature, hot and cold ensembles relax toward the same asymptotic state, and a macroscopic Fourier model produces an analogous exponential law in physical time \cite{LeonelThermo2016,Galia2017,HansenThermal2019}.

The purpose of this review is to discuss the following physical question: \emph{what ingredients are required for a chaotic mechanical model with a moving wall to behave, at the level of energy relaxation, like a system in contact with a finite-temperature reservoir?}

Following this question leads naturally from Fermi acceleration to diffusion, from diffusion to dissipation, from dissipation to a stationary kinetic temperature, and finally from collision counting to relaxation in physical time. The route provides a compact illustration of a broader theme in statistical physics: irreversible macroscopic behaviour may emerge only after a mechanical description has been coarse grained and expressed in terms of a restricted set of observables \cite{Reif1965,Pathria2011}.

\section{From a moving boundary to Fermi acceleration}
\label{sec:fermi}

\subsection{Collision dynamics with a moving wall}

Consider a classical point particle of mass $m$ moving freely inside a bounded two-dimensional domain. The boundary is allowed to vary in time and may be written in polar form as $R=R_b(\theta,t)$. 
Immediately after the $n$th impact, a convenient state is $(\theta_n,\alpha_n,V_n,t_n)$, where $\theta_n$ locates the collision point, $\alpha_n$ is the outgoing angle measured from the tangent, $V_n$ is the particle speed, and $t_n$ is the collision time. A static billiard requires only the geometrical collision variables; motion of the boundary therefore enlarges the map by introducing both the particle speed and the collision time.

Between collisions no force acts on the particle. The only non-trivial step is the reflection law. Let $\mathbf V_b$ be the instantaneous boundary velocity and define the relative particle velocity in the wall frame by
\begin{equation}
\mathbf V'_n=\mathbf V_n-\mathbf V_b.
\label{eq:relative_velocity}
\end{equation}
For a perfectly elastic impact, the tangential component of $\mathbf V'_n$ is preserved while the normal component changes sign. Transforming back to the laboratory frame then adds the wall motion to the reflected relative velocity. Consequently, even though the reflection is elastic in the instantaneous wall frame, in general
\begin{equation}
V_{n+1}\neq V_n,
\qquad
E_{n+1}\neq E_n,
\label{eq:energy_change}
\end{equation}
with $E_n=mV_n^2/2$.

\draftfigure[0.75\linewidth]{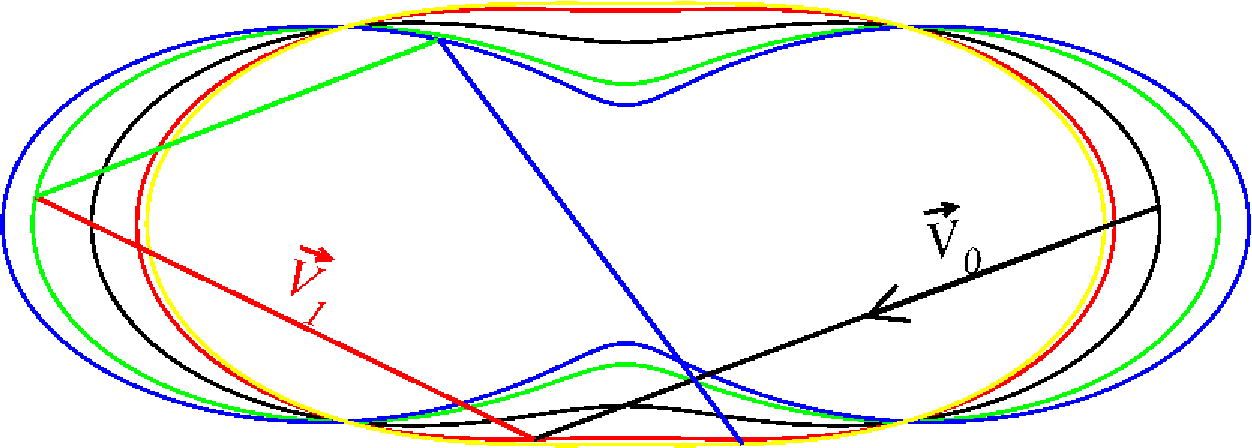}{Schematic collision sequence in a time-dependent billiard. In contrast with a static billiard, the wall occupies a different position at successive impacts. Elastic reflection in the instantaneous wall frame therefore produces a change of speed in the laboratory frame.}{fig:movingwall}

The geometry illustrated in Fig.~\ref{fig:movingwall} is enough to establish the main energetic idea. A moving wall performs work through impacts. Some collisions transfer energy to the particle, others remove it, and the long-time outcome is governed by the statistics of the collision sequence. Static chaotic billiards are a natural route to irregular sampling of wall phase. Loskutov, Ryabov and Akinshin argued that chaos in the corresponding static billiard supplies a sufficient mechanism for acceleration once a time-dependent perturbation is applied \cite{Loskutov2000}. The conjecture is not a statement of necessity: driven systems whose static limit is integrable can also accelerate when the perturbation destroys a separatrix and creates a stochastic layer, as demonstrated for the elliptical billiard \cite{Lenz2008}. More generally, modern work on non-autonomous billiards has clarified the roles of ergodicity, resonances, adiabatic invariants and stochastic energy growth \cite{Gelfreich2012,Itin2012}.

For the oval family used here, we take the time-dependent boundary to be
\begin{equation}
R(\theta,t)=1+\epsilon\left[1+\eta\cos(\omega t)\right]\cos(p\theta),
\label{eq:oval}
\end{equation}
where $\epsilon$ sets the static deformation, $\eta$ is the relative amplitude of the time-dependent modulation, $\omega$ is its angular frequency, and the integer $p$ fixes the angular symmetry. Unless stated otherwise, lengths are measured in units of the undeformed radius and time is rescaled so that $\omega=1$; in these dimensionless units the characteristic boundary-speed amplitude below is $A=\epsilon\eta$. In dimensional variables the corresponding speed scale contains the frequency factor $\omega$. In a chaotic component of the corresponding static billiard, the driven dynamics produces an extended regime in which the characteristic speed increases with the number of collisions \cite{LeonelOval2009}.

\subsection{Diffusion in velocity space}

Let $\Delta V_n=V_{n+1}-V_n$ denote an increment of the non-negative particle speed. For sufficiently rapid decorrelation, a simple diffusion picture assumes small systematic increments compared with their fluctuations:
\begin{equation}
\big|\langle \Delta V\rangle\big|\ll \sqrt{\langle(\Delta V)^2\rangle},
\qquad
\langle (\Delta V)^2\rangle\sim A^2,
\label{eq:increments}
\end{equation}
where $A$ is a characteristic wall-velocity scale. For a signed velocity component, independent kicks lead directly to a random-walk variance; an analogous scaling estimate for speed, away from its lower boundary and subject to suitable drift corrections, is
\begin{equation}
\langle (V-V_0)^2\rangle\sim A^2n,
\label{eq:diffusion_velocity}
\end{equation}
so that, when the initial-speed contribution is negligible, a root-mean-square scale grows as
\begin{equation}
V_{\rm rms}\sim A n^{1/2}.
\label{eq:sqrt_growth}
\end{equation}
The exponent $1/2$ is the familiar signature of normal diffusion. In the driven oval billiard an effective exponent close to this value is observed over a broad intermediate range, whereas correlations can generate crossovers to different transport regimes at longer times \cite{LeonelOval2009,Hansen2018}. Figure \ref{fig:fermi_growth} shows a plot of $V_{\rm rms}~vs.~n$ showing the regime of growth. A more systematic stochastic treatment starts from a one-step transition probability and the Chapman-Kolmogorov equation, making explicit the assumptions that enter a subsequent Fokker-Planck description \cite{Karlis2012}.

\draftfigure[0.75\linewidth]{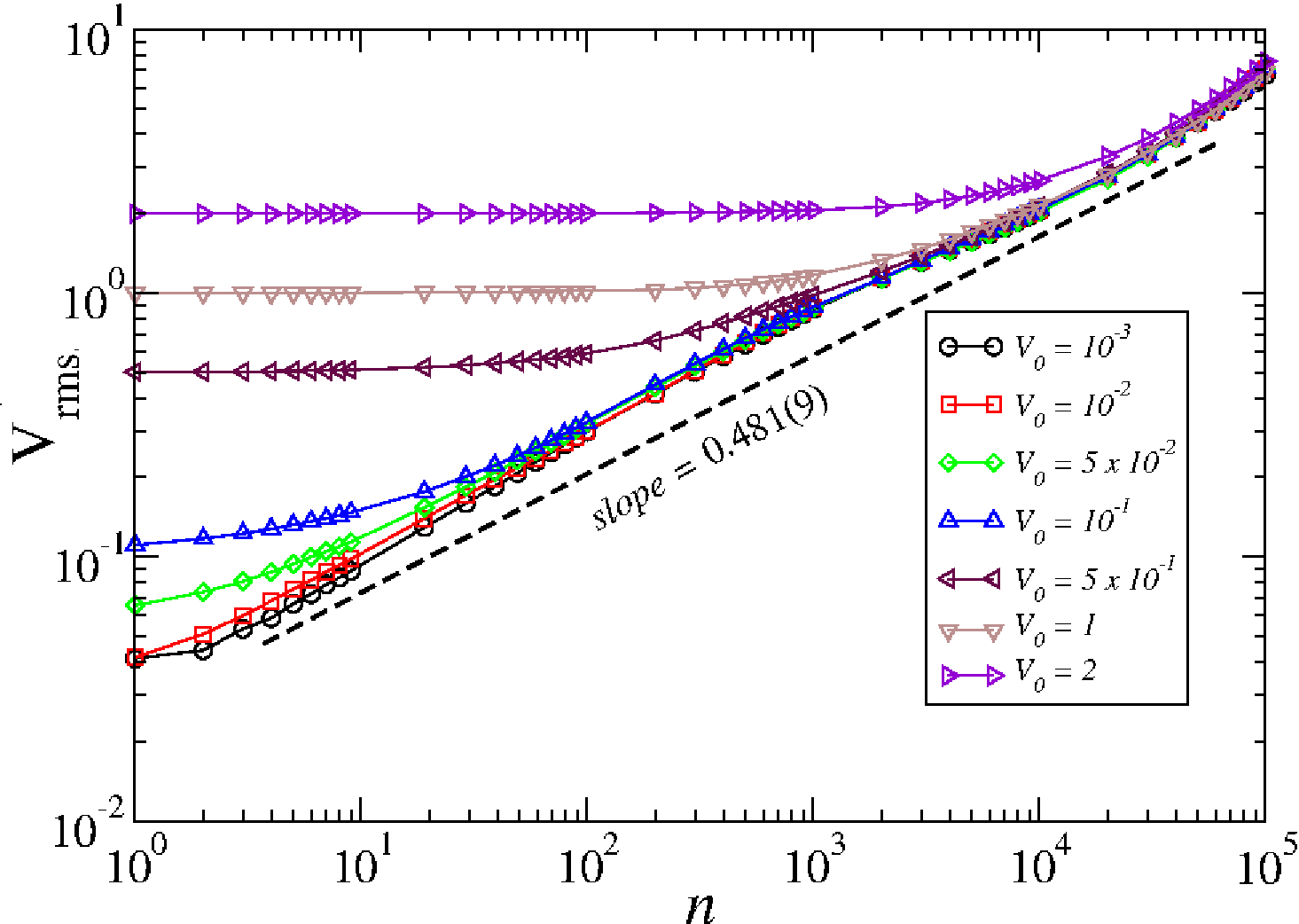}{Characteristic Fermi-acceleration curves for the time-dependent oval billiard. For sufficiently large initial speed the average velocity shows an initial plateau before crossing over to a growth regime; for small initial speed the accelerating regime is reached rapidly. The control parameters used were $\epsilon=0.08$, $p=3$ and $\eta=0.5$. The initial velocities are shown in the figure.}{fig:fermi_growth}

The important point for the present discussion is the absence of an upper energy scale. In the conservative driven model, diffusion is not opposed by any systematic loss mechanism. If the diffusive regime persists, the kinetic energy grows without bound:
\begin{equation}
\langle E\rangle
=\frac{m}{2}\langle V^2\rangle
\propto n.
\label{eq:energy_growth}
\end{equation}
A mechanical accelerator has therefore been constructed from nothing more than free flight and elastic collisions with a moving wall.

This is precisely where thermodynamics enters the story. If Eq.~\eqref{eq:energy_growth} is translated into a kinetic temperature, the model predicts indefinite heating. A finite-temperature wall should not do that. The accelerator contains fluctuating energy exchange, but it has no mechanism for enforcing a finite reservoir energy scale.

\section{Why the purely mechanical picture is incomplete}
\label{sec:incomplete}

A useful way to expose the missing physics is to imagine not one particle but a dilute ensemble of non-interacting particles inside the same moving container. Since the particles do not interact with one another, all energy exchange occurs at the wall. Introduce the kinetic-temperature convention
\begin{equation}
\frac{1}{2}m\langle V^2\rangle=k_BT.
\label{eq:kinetic_temperature}
\end{equation}
For a two-dimensional ensemble with no mean flow, this coincides with the usual kinetic equipartition convention. More generally, one should subtract the kinetic energy associated with any mean flow. Here Eq.~\eqref{eq:kinetic_temperature} defines an effective temperature based on the second velocity moment.

Combining Eqs.~\eqref{eq:energy_growth} and \eqref{eq:kinetic_temperature} gives $T_n\propto n$ within the simplest normal-diffusion regime. Thus the conservative time-dependent billiard behaves like an energy source of unlimited capacity. A wall at a prescribed temperature $T_b$, however, should instead impose a finite asymptotic energy scale. Standard thermodynamics predicts the direction of the relaxation without knowing any collision map: a cold gas should heat, a hot gas should cool, and both should approach the same bath temperature \cite{Reif1965,Pathria2011}.

The discrepancy shows that a prescribed conservative moving-wall motion is not a thermal-wall model. Boundary motion supplies fluctuations, but no irreversible channel returns energy from the particle motion to unresolved degrees of freedom of the wall. In reservoir models this missing channel can be introduced in different ways. A Langevin equation couples random forcing to friction, with their relation constrained near equilibrium by fluctuation-dissipation ideas \cite{Kubo1966}. Andersen dynamics uses stochastic velocity randomisation to sample a thermal ensemble \cite{Andersen1980}; Nos\'e and Hoover instead introduce deterministic extended dynamics designed to reproduce canonical statistics \cite{Nose1984,Hoover1985}. At gas--surface interfaces, kinetic models replace purely specular reflection by rules in which outgoing velocities partly or fully reflect the wall temperature \cite{Struchtrup2013,WuStruchtrup2017}. Random-billiard models make the same point geometrically: unresolved surface microstructure can generate stochastic reflection kernels with stationary Maxwell-Boltzmann-type distributions \cite{FeresYablonsky2004,CookFeres2012}.

The dissipative billiard considered next is simpler than these microscopic constructions. A constant restitution coefficient represents average energy transfer to the wall and, when needed, a random collision phase suppresses long correlations. This minimal phenomenology makes the energy balance analytically transparent and isolates the two ingredients that the present effective reservoir model requires:
\begin{equation}
{\text{fluctuating injection}
\quad + \quad
\text{irreversible loss}
\quad \Longrightarrow \quad
\text{finite stationary energy scale}.}
\label{eq:reservoir_principle}
\nonumber
\end{equation}

\section{A fluctuating and dissipative wall}
\label{sec:dissipative}

\subsection{Inelastic reflection in the wall frame}

Let $\mathbf T$ and $\mathbf N$ denote the unit tangent and normal at the impact point. At collision $n+1$, the incoming and outgoing relative velocities must be evaluated with the \emph{same} instantaneous wall velocity: $\mathbf V'_{n+1,\mathrm{in}}=\mathbf V_n-\mathbf V_b(t_{n+1})$ and $\mathbf V'_{n+1,\mathrm{out}}=\mathbf V_{n+1}-\mathbf V_b(t_{n+1})$, where the bold velocities denote vectors. An inelastic law preserves the tangential relative component while reducing and reversing the normal component:
\begin{align}
\mathbf V'_{n+1,\mathrm{out}}\!\cdot\mathbf T_{n+1}
&=\mathbf V'_{n+1,\mathrm{in}}\!\cdot\mathbf T_{n+1},
\label{eq:tangent_rule}\\
\mathbf V'_{n+1,\mathrm{out}}\!\cdot\mathbf N_{n+1}
&=-\gamma\,\mathbf V'_{n+1,\mathrm{in}}\!\cdot\mathbf N_{n+1},
\qquad 0\leq\gamma\leq1.
\label{eq:normal_rule}
\end{align}
The elastic limit is $\gamma=1$. For $\gamma<1$, a fraction of the normal relative kinetic energy is lost at every collision. Returning to the laboratory frame adds the instantaneous wall velocity, so the same collision still contains a fluctuating energy-injection term. This competition has been studied numerically in the oval billiard and in other driven billiards, where even weak dissipation suppresses unbounded Fermi acceleration \cite{OliveiraLeonel2010,LeonelBunimovich2010}.

The short-time behaviour remains instructive. When $1-\gamma$ is small, many impacts are required before the cumulative loss becomes comparable to the stochastic injection. The early motion therefore resembles the conservative random walk,
\begin{equation}
V_{\rm rms}\propto n^{1/2},
\label{eq:early_diffusive}
\end{equation}
but the growth bends over after a characteristic number of collisions and approaches a plateau.

\draftfigure[0.45\linewidth]{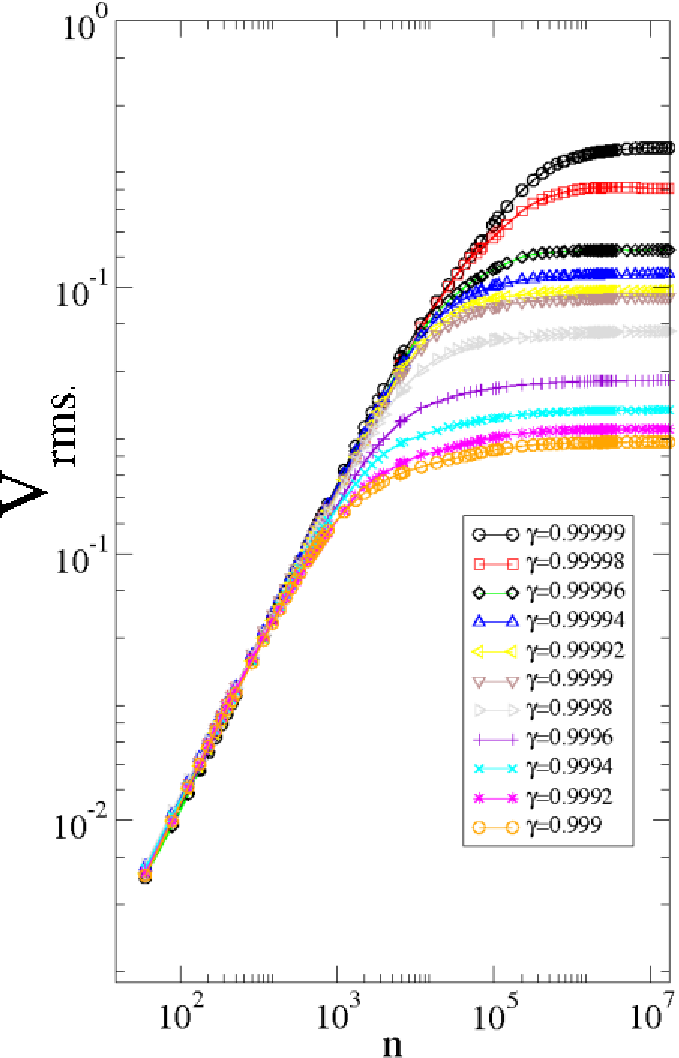}{Suppression of Fermi acceleration by inelastic collisions. For weak dissipation the early evolution follows an approximately common diffusive growth, but every finite restitution loss ultimately produces a stationary velocity plateau. The control parameters used in the construction of the figure were $p=2$, $\epsilon= 0.4$ and $\eta=0.001$.}{fig:suppression}

The qualitative sequence in Fig.~\ref{fig:suppression} is
\begin{equation}
\text{diffusive growth}
\longrightarrow
\text{crossover}
\longrightarrow
\text{stationary state}.
\label{eq:growth_cross_sat}
\nonumber
\end{equation}
The essential question is whether this sequence can be obtained analytically without solving the full four-dimensional deterministic map.

\subsection{The decorrelated-impact approximation}

In a strongly chaotic regime, collision position, incidence angle and wall phase can lose memory sufficiently rapidly that an averaged description becomes useful. Denote the wall-velocity scale by $A=\eta\epsilon$, using the notation of the driven oval boundary. Averaging the squared post-collision speed under the decorrelation approximation gives \cite{LeonelThermo2016}
\begin{equation}
{S_{n+1}
=\frac{1+\gamma^2}{2}S_n
+\frac{(1+\gamma)^2}{8}A^2,
}
\qquad {\rm with}\qquad
S_n\equiv\langle V_n^2\rangle.
\label{eq:second_moment_recurrence}
\end{equation}
This one-line recurrence contains the essential heating-cooling structure of the effective microscopic model. Here $S_n$ denotes an ensemble moment at fixed collision index $n$.

It is useful to write
\begin{equation}
a=\frac{1+\gamma^2}{2},
\qquad
b=\frac{(1+\gamma)^2}{8}A^2,
\label{eq:a_b}
\end{equation}
so that $S_{n+1}=aS_n+b$. For $\gamma<1$, one has $0<a<1$, and the exact solution is
\begin{equation}
S_n=S_\infty+(S_0-S_\infty)a^n,
\label{eq:S_solution}
\end{equation}
with
\begin{equation}
S_\infty=\frac{b}{1-a}
=\frac{(1+\gamma)A^2}{4(1-\gamma)}.
\label{eq:S_infty}
\end{equation}
The plateau is therefore not an arbitrary fitting scale: it is the fixed point of the energy-balance recurrence.

Near the elastic limit, let $\rho=1-\gamma\ll1$. Then
\begin{equation}
V_{\rm rms,\infty}=\sqrt{S_\infty}
\sim \frac{A}{\sqrt{2}}\rho^{-1/2},
\label{eq:Vsat_scaling}
\end{equation}
whereas the relaxation factor satisfies
\begin{equation}
a^n\simeq\exp\left[-\frac{1-\gamma^2}{2}n\right].
\label{eq:a_exp}
\end{equation}
The corresponding collision-number relaxation scale diverges as
\begin{equation}
\tau_n
=-\frac{1}{\ln[(1+\gamma^2)/2]}
\simeq\frac{2}{1-\gamma^2}
\sim\rho^{-1}.
\label{eq:tau_n}
\end{equation}
Thus removing dissipation eliminates both a finite energy scale and a finite relaxation scale.

\subsection{Stationarity as injection-loss balance}

The recurrence \eqref{eq:second_moment_recurrence} can be rearranged as
\begin{equation}
S_{n+1}-S_n
=-\frac{1-\gamma^2}{2}S_n
+\frac{(1+\gamma)^2}{8}A^2.
\label{eq:balance}
\end{equation}
The first term is an average loss proportional to the current kinetic scale, whereas the second is the mean injection produced by the moving boundary. Stationarity is reached when the two contributions cancel. This interpretation is more informative than simply saying that the velocity ``saturates'': the chaotic motion need not disappear, and the stationary state remains dynamically active, sustained by continuous energy exchange:
\begin{equation}
{\text{mean energy injected per collision}
=
\text{mean energy dissipated per collision}.}
\label{eq:balance_box}
\nonumber
\end{equation}

Comparable balance ideas are ubiquitous in driven granular systems, where inelastic collisions remove kinetic energy while external forcing maintains a statistically stationary state \cite{BrilliantovPoschel2004}. The present billiard is much simpler, but its simplicity makes the mechanism unusually transparent.

\section{From stationary velocity to kinetic temperature}
\label{sec:temperature}

Equation~\eqref{eq:S_infty} becomes thermodynamically suggestive when combined with the kinetic-temperature definition \eqref{eq:kinetic_temperature}. We obtain
\begin{equation}
{T_\infty
=\frac{mA^2}{8k_B}
\frac{1+\gamma}{1-\gamma}.
}
\label{eq:T_infty}
\end{equation}
This expression is a central bridge between the billiard dynamics and the effective reservoir picture \cite{LeonelThermo2016}. It makes the respective roles of forcing and dissipation explicit: increasing $A$ raises the effective temperature quadratically, whereas increasing $\gamma$ reduces the loss per impact and therefore raises the stationary temperature. In the conservative limit,
\begin{equation}
\gamma\rightarrow1^-
\quad\Longrightarrow\quad
T_\infty\rightarrow\infty,
\label{eq:Tdiv}
\end{equation}
which recovers the absence of a stationary scale in the Fermi accelerator.

The result should be interpreted carefully. Equation \eqref{eq:T_infty} is the temperature associated with the second moment of an effective stochastic-dissipative collision model. If the wall is to represent a reservoir at prescribed temperature $T_b$, thermodynamic consistency requires
\begin{equation}
T_\infty=T_b.
\label{eq:matching_temp}
\end{equation}
Within the model this becomes a calibration relation between forcing, dissipation and bath temperature:
\begin{equation}
\frac{1+\gamma}{1-\gamma}
=\frac{8k_BT_b}{mA^2}.
\label{eq:calibration}
\end{equation}
Equation~\eqref{eq:calibration} should be read as a consistency condition within the coarse-grained model. Once the fluctuation amplitude is fixed, the mean dissipation must be calibrated if the stationary kinetic temperature is to match a prescribed bath temperature. Because $0\leq\gamma<1$, this calibration is possible at fixed $A$ only for $T_b\geq mA^2/(8k_B)$ in this particular simplified model.

This is closely related in spirit to thermostat constructions in molecular dynamics. Andersen's stochastic thermostat, Nos\'e's extended-system method and Hoover's dynamical reformulation differ substantially in implementation, but each introduces a mechanism by which a mechanical subsystem exchanges energy with an environment while approaching a prescribed statistical state \cite{Andersen1980,Nose1984,Hoover1985}. Random billiards with wall temperature provide an even closer geometrical analogy: stochastic collision kernels can generate invariant distributions tied to a boundary temperature \cite{CookFeres2012,KhaninYarmola2013}. The oval billiard model is not equivalent to these thermostats, but it belongs to the same broad conceptual family of effective reservoir couplings.

\section{Heating and cooling toward the same stationary state}
\label{sec:heatingcooling}

A useful feature of Eq.~\eqref{eq:S_solution} is that the same dynamical rule describes both heating and cooling. Using Eq.~\eqref{eq:kinetic_temperature}, the exact collision-number evolution is
\begin{equation}
{T_n
=T_\infty+(T_0-T_\infty)
\left(\frac{1+\gamma^2}{2}\right)^n.
}
\label{eq:Tn_exact}
\end{equation}
If $T_0<T_\infty$, the ensemble heats. If $T_0>T_\infty$, it cools. Both branches approach the same fixed point.

For weak dissipation the relaxation may be written in the familiar exponential form
\begin{equation}
T_n-T_\infty
\propto e^{-n/\tau_n},
\label{eq:Tn_exp}
\end{equation}
with $\tau_n$ given by Eq.~\eqref{eq:tau_n}. The sign of $T_0-T_\infty$ determines the direction of energy flow, while the same relaxation factor controls the approach from either side. This is the qualitative structure expected for relaxation toward a single reservoir-controlled state.

Numerical studies of stochastic dissipative billiards show the corresponding evolution of velocity moments and distributions, including relaxation from both high- and low-energy initial ensembles toward the same asymptotic scale \cite{Galia2017,HansenThermal2019}. The dynamics can display an initial diffusive plateau or growth regime, an intermediate exponential relaxation, and a final stationary stage. These multiple stages are useful reminders that ``thermalisation time'' is not always captured by a single short-time fit.

The full probability distribution evolves as well, as illustrated in Fig.~\ref{fig:distribution}. Starting from a narrow ensemble, the early motion may exhibit approximately symmetric broadening. Once the lower kinematic boundary becomes relevant, that symmetry is lost, and the distribution eventually approaches a stationary non-Gaussian form with a finite second moment. A convergence of a second moment is enough to define a kinetic temperature, but not enough to establish canonical equilibrium.

\draftfigure[0.7\linewidth]{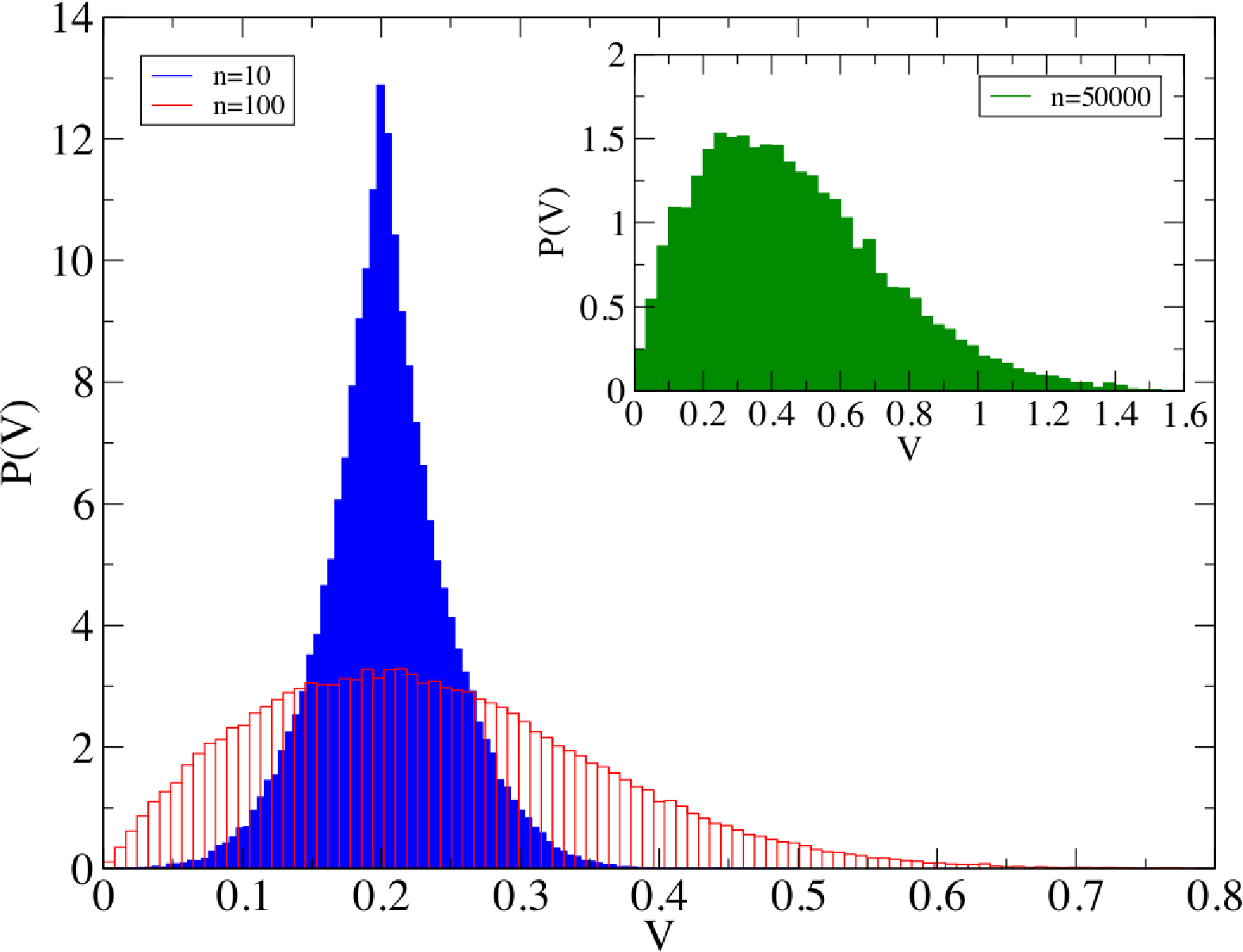}{Evolution of the velocity distribution in the stochastic dissipative oval billiard. A narrow early-time ensemble broadens under fluctuating kicks; at long times the distribution becomes stationary and asymmetric with a decaying high-velocity tail. The control parameters used were $p=2$, $\eta=0.02$, $\gamma=0.999$ for the initial velocity $V_0=0.2$.}{fig:distribution}

This distinction is familiar from granular gases, where dissipation can generate stationary or cooling states with non-Maxwellian velocity statistics even though temperature-like quantities remain useful \cite{BrilliantovPoschel2004}. The existence of a stationary measure and the approach toward it are separate questions from the detailed form of that measure. Random-billiard thermostat models make this distinction mathematically explicit by characterising invariant distributions and convergence properties of the associated Markov chains \cite{CookFeres2012,KhaninYarmola2013}.

\section{What macroscopic thermodynamics expects}
\label{sec:fourier}

The microscopic recurrence becomes more informative when placed beside a deliberately simple macroscopic description. Suppose the particle gas exchanges heat with a wall held at temperature $T_b$. A one-dimensional Fourier picture writes
\begin{equation}
\frac{dQ}{dt}
=-\kappa\ell\frac{\partial T}{\partial x},
\label{eq:fourier_law}
\end{equation}
where $\kappa$ is an effective thermal conductivity and $\ell$ is the effective wall length participating in the transfer. Fourier's heat-conduction theory provides the classical macroscopic framework for such diffusive thermal relaxation \cite{Fourier1878,Narasimhan1999}.

For the time-dependent oval billiard, the region swept by the boundary defines a natural collision-zone thickness, as shown in Fig.~\ref{fig:heatflow}. With the boundary in Eq.~\eqref{eq:oval}, an rms estimate of the radial width swept over one cycle gives
\begin{equation}
\Delta x=\sqrt{2}\,A,
\label{eq:delta_x}
\end{equation}
while the simplified angularly integrated geometric factor is $\ell=\int_0^{2\pi}R(\theta,t)\,d\theta=2\pi$ in the dimensionless model.Approximating the temperature gradient by
\begin{equation}
\frac{\partial T}{\partial x}\simeq\frac{T-T_b}{\Delta x},
\label{eq:grad_approx}
\end{equation}
and writing $dQ=c\,dT$ gives
\begin{equation}
c\frac{dT}{dt}
=-\frac{\kappa\ell}{\Delta x}(T-T_b).
\label{eq:heat_ode}
\end{equation}
Defining
\begin{equation}
\tau_T=\frac{c\Delta x}{\kappa\ell},
\label{eq:tauT}
\end{equation}
we obtain
\begin{equation}
{T(t)=T_b+(T_0-T_b)e^{-t/\tau_T}.
}
\label{eq:Fourier_solution}
\end{equation}

\draftfigure[0.7\linewidth]{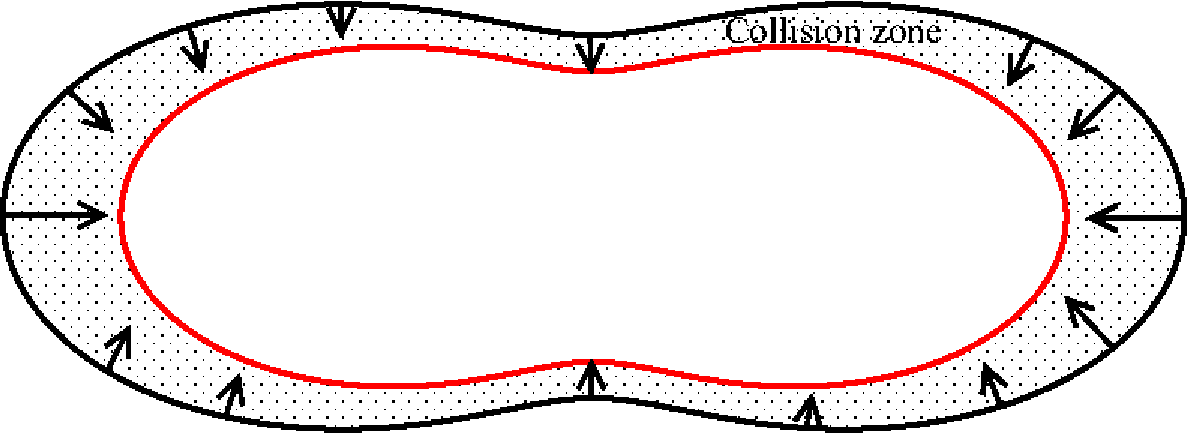}{Macroscopic heat-transfer picture. When the particle gas is colder than the wall, heat flows from the reservoir into the gas; when the gas is hotter, the direction reverses. In the simplified model the same bath temperature is approached from either side.}{fig:heatflow}

Equations~\eqref{eq:Tn_exp} and \eqref{eq:Fourier_solution} display a useful structural parallel:
\begin{equation}
{T_n-T_\infty\propto e^{-n/\tau_n},
\qquad
T(t)-T_b\propto e^{-t/\tau_T}.
}
\label{eq:parallel_relax}
\end{equation}
The first law emerges from a collision map after coarse-grained averaging; the second follows from a lumped Fourier-inspired heat-transfer equation. Their common exponential structure is the bridge between the two descriptions. The variables $n$ and $t$, however, are not interchangeable, so the similarity of the relaxation laws should not be mistaken for an identity of the underlying dynamics.

The macroscopic treatment also clarifies the short-time heating branch. If $T_0\ll T_b$, expansion of Eq.~\eqref{eq:Fourier_solution} gives
\begin{equation}
T(t)\simeq T_0+(T_b-T_0)\frac{t}{\tau_T}.
\label{eq:short_time_T}
\end{equation}
For an initially cold ensemble, over a short interval where the linear term dominates the initial offset, $T\propto t$, so Eq.~\eqref{eq:kinetic_temperature} implies
\begin{equation}
V_{\rm rms}\propto t^{1/2}.
\label{eq:Vrms_t}
\end{equation}
Thus the square-root law appears from both a diffusion argument in collision number and a heat-transfer argument in physical time. Their numerical exponents coincide, but the clocks and prefactors encode different physics.

\section{Two clocks: collision number is not physical time}
\label{sec:clocks}

Billiard dynamics is naturally formulated as a map from one impact to the next, so its independent variable is the collision number $n$. Thermodynamic relaxation, by contrast, is expressed in physical time. The distinction is essential rather than notational. A faster particle makes more collisions per unit time than a slower one, so the conversion factor changes during heating or cooling.

The exact elapsed time after $n$ collisions is
\begin{equation}
{t_n=\sum_{i=0}^{n-1}\frac{d_i}{V_i},
}
\label{eq:tn_sum}
\end{equation}
where $d_i$ is the free-flight distance between two successive impacts. For a chaotic ensemble one may introduce a mean free-flight distance $\bar d$ and write, in a slowly varying approximation,
\begin{equation}
t_n\simeq\bar d\int_0^n\frac{dn'}{V(n')}.
\label{eq:tn_integral}
\end{equation}
During a transient, $V(n')$ changes substantially and the integral must be retained. There is no constant proportionality between $t$ and $n$.

After the stationary regime is reached, the collision-by-collision average interval may become constant, provided its mean is finite. A further mean-field estimate replaces the distribution of flight times by a typical flight distance divided by a typical speed:
\begin{equation}
t_n\simeq n\left\langle\frac{d}{V}\right\rangle_{\rm st}\approx\frac{\bar d}{V_{\rm rms,\infty}}n.
\label{eq:tn_stationary}
\end{equation}
With Eq.~\eqref{eq:Vsat_scaling},
\begin{equation}
t_n\propto\frac{\bar d\sqrt{1-\gamma}}{A}\,n,
\label{eq:two_clock_scaling}
\end{equation}
near the elastic limit. We emphasize this is a scaling estimate, not an exact identity between ensemble-averaged inverse speed and inverse rms speed.

This observation removes a common ambiguity in comparisons between discrete maps and continuous-time relaxation. Fitting a temperature curve against collision number and immediately calling the resulting scale a ``time hides the changing collision frequency and the distinction between $\langle 1/V\rangle$ and $1/\sqrt{\langle V^2\rangle}$. The correct conversion is state dependent during the transient. The distinction matters especially near the conservative limit because the stationary speed diverges while the collision-number relaxation scale also diverges. These competing trends affect the physical-time relaxation in a way that cannot be inferred from $\tau_n$ alone.

The issue is not unique to billiards. Many coarse-grained models use an event count, Monte Carlo sweep, or collision index as a natural dynamical variable, whereas experimentally measured relaxation occurs in laboratory time. Establishing a physically meaningful map between the two clocks is part of the modelling problem rather than a bookkeeping detail.

\section{Microscopic dynamics versus Fourier heat transfer}
\label{sec:comparison}

We can now place the two descriptions side by side. The macroscopic model contains no explicit chaotic dynamics. It assumes a temperature field, an effective conductivity and a fixed reservoir temperature. The microscopic model contains no phenomenological heat equation. It follows impact-to-impact changes of particle velocity under a moving, fluctuating and dissipative boundary. Yet both produce the same qualitative heating-cooling structure:
\begin{equation}
\begin{array}{lll}
T_0<T_* &\Longrightarrow& \text{heating},\\[1mm]
T_0>T_* &\Longrightarrow& \text{cooling},\\[1mm]
T_0=T_* &\Longrightarrow& \text{stationarity},
\end{array}
\label{eq:branches}
\end{equation}
where $T_*=T_b$ macroscopically and $T_*=T_\infty$ microscopically.

Matching the two requires Eq.~\eqref{eq:matching_temp}. Once this is imposed, the two models describe the same asymptotic energy scale at different levels of resolution; equality of the characteristic relaxation times would require an additional relation between the microscopic collision statistics and the effective macroscopic transport parameters. The comparison can be summarised as
\begin{equation}
\boxed{\begin{array}{ccc}
\text{macroscopic} && \text{microscopic}\\
T_b && T_\infty\\
\tau_T && \tau_n\\
\text{Fourier flux} && \text{collision energy balance}\\
\text{physical time }t && \text{collision number }n.
\end{array}}
\label{eq:comparison_table_eq}
\nonumber
\end{equation}

The microscopic recurrence \eqref{eq:second_moment_recurrence} derives a stationary second moment and a relaxation factor from an averaged collision rule. The macroscopic model derives an exponential approach from energy conservation plus a constitutive heat-flux relation. Agreement of the asymptotic temperature and of the qualitative relaxation law does not mean that either description has been rigorously derived from the other. It means only that, within their stated assumptions, they provide mutually consistent coarse-grained descriptions of the same energy exchange \cite{LeonelThermo2016}.

Random-billiard theory provides useful context for this interpretation. Feres and Yablonsky showed how deterministic microgeometry can motivate stochastic gas-surface rules and stationary angular distributions \cite{FeresYablonsky2004}. Cook and Feres developed random billiards with wall temperature whose invariant measures include familiar equilibrium distributions \cite{CookFeres2012}. Chumley, Cook and Feres used related random mechanical models to explore the emergence of thermodynamic behaviour and heat-engine-like steady states \cite{Chumley2013}. These constructions reinforce the idea that a collision model can sit naturally between deterministic mechanics and thermodynamics when unresolved boundary degrees of freedom are represented statistically.

The converse caution is equally important. Fourier's law is a macroscopic constitutive relation. Its success in the present comparison rests on the existence of a slowly varying temperature-like variable and a coarse-grained heat flux. In strongly non-equilibrium regimes, in rarefied gases close to walls, or when the velocity distribution is far from local equilibrium, kinetic boundary conditions and higher-order descriptions may be required \cite{Struchtrup2013,WuStruchtrup2017}.

\section{The elastic limit and the loss of a thermal scale}
\label{sec:elasticlimit}

The conservative limit deserves separate attention because it brings the mechanical and thermodynamic pictures into sharp contrast. Let
\begin{equation}
\rho=1-\gamma.
\label{eq:rho}
\end{equation}
For weak dissipation, Eqs.~\eqref{eq:T_infty} and \eqref{eq:tau_n} imply
\begin{equation}
T_\infty\propto A^2\rho^{-1},
\qquad
\tau_n\propto\rho^{-1}.
\label{eq:divergences}
\end{equation}
Thus the stationary energy scale and the number of collisions required to reach it diverge together as $\rho\rightarrow0^+$. Within the decorrelated-impact approximation, any fixed non-zero $\rho$ gives a stationary second moment; at $\rho=0$ the corresponding accelerating regime has no finite energy plateau.

For negligible initial speed, this singular behaviour can be phrased in the scaling form
\begin{equation}
V_{\rm rms}(n,\rho,A)
=A\rho^{-1/2}\,\mathcal G(n\rho),
\label{eq:master_scaling}
\end{equation}
with
\begin{equation}
\mathcal G(x)\propto x^{1/2}
\quad(x\ll1),
\qquad
\mathcal G(x)\rightarrow\text{const.}
\quad(x\gg1).
\label{eq:G_asympt}
\end{equation}
The early branch reproduces the conservative diffusive law $V_{\rm rms}\sim An^{1/2}$; the late branch gives $V_{\rm rms,\infty}\sim A\rho^{-1/2}$, as shown in Fig.~\ref{fig:divergence}. In the thermal language the same structure says that arbitrarily weak dissipation eventually creates a finite temperature, but the temperature and relaxation scale become arbitrarily large as the conservative accelerator is approached.

\draftfigure[0.65\linewidth]{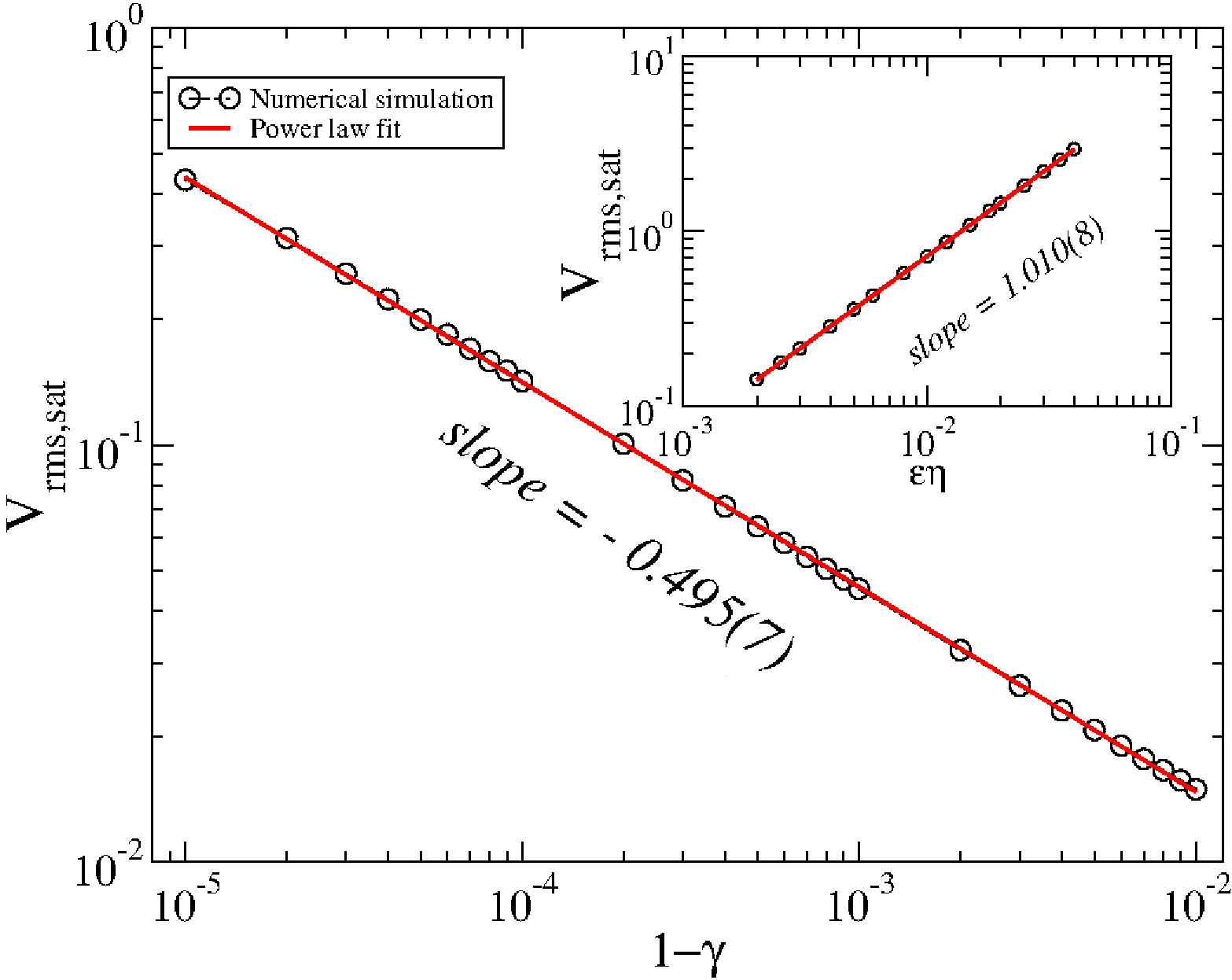}{Weak-dissipation scaling of the stationary root-mean-square speed. The main plot shows its dependence on $\rho=1-\gamma$, with the measured exponent $\alpha_1=-0.495(7)$; the inset shows the dependence on forcing amplitude $A$, with the measured exponent $\alpha_2=1.010(8)$. The theoretical predictions are $\alpha_1=-1/2$ and $\alpha_2=1$.}{fig:divergence}

The thermodynamic message is simpler: within this effective model, dissipation sets both the finite energy scale and the relaxation scale. As that coupling is removed, both scales diverge and the finite-temperature reservoir interpretation breaks down.

\section{What does ``thermalisation'' mean here?}
\label{sec:meaning}

The word thermalisation can carry several meanings, and the billiard model benefits from keeping them separate.

First, there is \emph{energetic stationarity}: the second velocity moment converges to a finite value. This property follows directly from Eq.~\eqref{eq:second_moment_recurrence} for $\gamma<1$. It is enough to define $T_\infty$ through Eq.~\eqref{eq:kinetic_temperature}.

Second, there is \emph{probability-level stationarity}: the full velocity distribution approaches a time-independent shape. Numerical studies support such convergence for stochastic dissipative time-dependent billiards, with an asymptotic distribution that has a finite characteristic scale and convergent low-order moments \cite{Galia2017,HansenThermal2019}.

Third, there is \emph{canonical equilibrium}: the stationary distribution is specifically Maxwell-Boltzmann and all relevant degrees of freedom are consistent with a canonical ensemble. The effective billiard model does not establish this stronger statement. The particles do not collide with one another, the wall is represented through a random phase and a constant restitution coefficient, and the stationary distribution is generated by the collision dynamics rather than derived from an atomistic reservoir. 

This hierarchy is useful beyond billiards. Driven granular gases and thermostatted dynamical systems, for example, can possess temperature-like observables without being equilibrium systems \cite{BrilliantovPoschel2004,Andersen1980,Nose1984,Hoover1985}. Conversely, mathematically constructed stochastic billiards can be designed so that a genuine Maxwellian stationary law is built into the wall interaction \cite{CookFeres2012,KhaninYarmola2013}. The oval billiard occupies an intermediate position: it is intended to expose the mechanism of stationary energy balance rather than to reproduce the full microscopic statistics of a solid surface.

The analogy with fluctuation-dissipation physics should be handled in the same spirit. A random phase and inelastic loss mimic, respectively, fluctuating forcing and dissipation, but Eq.~\eqref{eq:calibration} is not an instance of the Kubo fluctuation-dissipation theorem. Kubo's result connects linear response to equilibrium fluctuations under specific statistical assumptions \cite{Kubo1966}. Here we have instead a phenomenological calibration in a driven collision model.

\section{Beyond the minimal model}
\label{sec:beyond}

The simplicity of the stochastic inelastic billiard is both its main advantage and its main limitation. Several extensions point toward a broader research programme.

\subsection{More realistic wall interactions}

A constant restitution coefficient compresses complicated microscopic energy transfer into a single number. Real gas-surface interactions depend on incidence angle, incoming speed, surface roughness and wall temperature. Maxwell-type accommodation models and their velocity-dependent extensions explicitly distinguish specular and diffuse components of reflection \cite{Struchtrup2013,WuStruchtrup2017}. Random billiards with microstructure go further by deriving transition kernels from mechanical surface models and recovering stationary distributions such as the Knudsen cosine law or Maxwell-Boltzmann statistics under suitable conditions \cite{FeresYablonsky2004,CookFeres2012}. Embedding time dependence into such models could connect Fermi acceleration more directly to kinetic gas-surface theory.

\subsection{Thermostats and deterministic reservoirs}

The present effective reservoir uses randomness explicitly, which raises the complementary question of how much of the same thermal behaviour can be generated deterministically. Nos\'e-Hoover dynamics is the canonical example of a deterministic thermostat, but billiard-specific constructions also exist in which collisions exchange energy with thermostat degrees of freedom \cite{Nose1984,Hoover1985,KhaninYarmola2013}. Comparing deterministic and stochastic reservoir models is particularly interesting when the underlying billiard is chaotic, because deterministic mixing may provide part of the decorrelation that the random phase supplies by construction.

\subsection{Correlations and anomalous acceleration}

The recurrence \eqref{eq:second_moment_recurrence} assumes sufficient decorrelation. Long-lived correlations, intermittent phase-space structures and resonant passages can invalidate the simple normal-diffusion picture. Driven elliptical and rectangular billiards show that time-dependent integrable geometries can develop acceleration through stochastic layers or resonance crossings \cite{Lenz2008,Itin2012}. General non-autonomous billiards can display energy growth laws controlled by ergodicity and adiabatic invariants \cite{Gelfreich2012}. Even in the oval billiard, a crossover from normal to faster diffusion has been identified at long times \cite{Hansen2018}. Understanding how such correlations modify the effective temperature and relaxation laws is a natural next step.

\subsection{Many-particle effects}

The dilute-gas assumption removes particle-particle collisions. In a real gas, intermolecular scattering redistributes energy and drives the velocity distribution toward local equilibrium. Adding such collisions would introduce a second route to decorrelation and could separate two relaxation processes: equilibration within the gas and energy exchange with the wall. Kinetic theory and granular-gas methods offer established languages for this problem \cite{BrilliantovPoschel2004,Pathria2011}. The minimal billiard model is therefore best viewed as the wall-coupling limit of a larger nonequilibrium system.

\subsection{From one bath to heat transport}

A single thermal wall establishes a stationary temperature. Two reservoirs at different temperatures introduce a genuine nonequilibrium steady state with sustained heat current and entropy production. Random-billiard models have been used to construct heat-engine-like dynamics and thermostat-driven steady states \cite{Chumley2013,KhaninYarmola2013}. Time-dependent billiards could add an extra control knob: the wall motion can inject mechanical work while temperature differences drive heat flow. Such systems would offer an accessible setting in which to separate work, heat and dissipation at the level of individual collision events.

\section{Outlook and concluding remarks}
\label{sec:outlook}

A particle bouncing from a moving wall looks at first like a textbook mechanics problem. Pursued far enough, however, the problem leads directly into nonlinear dynamics, diffusion and thermodynamics. The reason is that a moving boundary does more than reflect trajectories: it mediates energy exchange. When the collision sequence decorrelates sufficiently, the exchanges can be described statistically and the characteristic speed can grow without bound. This is the essence of Fermi acceleration.

The same mechanism also makes clear why a conservative moving wall cannot represent a finite-temperature reservoir. It supplies fluctuations without a compensating energy sink. Introducing inelasticity changes the qualitative long-time behaviour. The early motion retains the diffusive signature of the accelerator, but the accumulation of losses creates a crossover and a stationary kinetic-energy scale. In the decorrelated-impact approximation the entire process is encoded by the recurrence
\begin{equation}
\langle V_{n+1}^2\rangle
=\frac{1+\gamma^2}{2}\langle V_n^2\rangle
+\frac{(1+\gamma)^2}{8}A^2,
\nonumber
\end{equation}
whose fixed point yields
\begin{equation}
{T_\infty
=\frac{mA^2}{8k_B}
\frac{1+\gamma}{1-\gamma}.}
\nonumber
\end{equation}
The equation has a simple interpretation: forcing determines the strength of the fluctuations, dissipation sets the ability of the boundary to remove energy, and their balance defines the stationary kinetic temperature.

The microscopic relaxation law
\begin{equation}
T_n-T_\infty\propto e^{-n/\tau_n},
\nonumber
\end{equation}
then mirrors the macroscopic Fourier form
\begin{equation}
T(t)-T_b\propto e^{-t/\tau_T}.
\nonumber
\end{equation}
The parallel is physically appealing but should not obscure the difference between the two clocks. Collision number is an event count; physical time depends on the speed and free-flight distance. Their relation is nonlinear during transient heating and cooling and becomes approximately linear only after stationarity is reached.

The resulting picture provides a useful, explicitly approximate bridge between levels of description. At one end sits a deterministic or stochastic collision process with explicit energy kicks. At the other sits a macroscopic relaxation law with a temperature and a heat flux. Between them lies coarse graining: correlations are discarded, unresolved wall dynamics is represented statistically, and a second velocity moment is promoted to a temperature-like observable.

This bridge is useful because each side illuminates the other. Nonlinear dynamics identifies how unbounded acceleration emerges and how dissipation suppresses it. Thermodynamics explains why fluctuations alone are insufficient for a reservoir and why a stationary state requires compensating loss. Random-billiard theory, kinetic gas-surface models and thermostat dynamics show that this connection is part of a wider family of approaches in which mechanical collision rules generate statistical behaviour \cite{CookFeres2012,Chumley2013,WuStruchtrup2017}.

The central lesson can be condensed into one sequence: moving wall $\rightarrow$ irregular energy kicks $\rightarrow$ Fermi acceleration $\rightarrow$ dissipation $+$ fluctuations $\rightarrow$ stationary energy balance $\rightarrow$ kinetic temperature $\rightarrow$ thermal relaxation. A ball colliding with a wall thus provides a route from elementary mechanics to a non-trivial question at the heart of statistical physics: how does a finite thermal scale emerge from microscopic dynamics?

\section*{Acknowledgements}
E.D.L. acknowledges support from the Brazilian agencies CNPq (No. 304398/2023-3) and FAPESP (2025/14544-0).

\section*{Disclosure statement}
No potential conflict of interest was reported by the author.

\end{document}